\documentclass[%
 aip,
 amsmath,amssymb, floatfix,
 reprint,%
]{revtex4-1}

\usepackage{graphicx}
\usepackage{dcolumn}
\usepackage{bm}
\usepackage{siunitx}
\usepackage{amsmath}
\usepackage{environ}
\usepackage[utf8]{inputenc}
\usepackage[T1]{fontenc}
\usepackage{mathptmx}
\usepackage{etoolbox}
\usepackage{physics}

\makeatletter
\def\@email#1#2{%
 \endgroup
 \patchcmd{\titleblock@produce}
  {\frontmatter@RRAPformat}
  {\frontmatter@RRAPformat{\produce@RRAP{*#1\href{mailto:#2}{#2}}}\frontmatter@RRAPformat}
  {}{}
}%
\makeatother
\begin{document}

\preprint{AIP/123-QED}

\title[Carrier confinement and alloy disorder exacerbate Auger-Meitner recombination in AlGaN ultraviolet light-emitting diodes]{Carrier confinement and alloy disorder exacerbate Auger-Meitner recombination in AlGaN ultraviolet light-emitting diodes}
\author{Nick Pant}
\email{nickpant@umich.edu, kioup@umich.edu}
\affiliation{ 
Applied Physics Program, University of Michigan, Ann Arbor, MI, USA 48109} %
\affiliation{ 
Department of Materials Science \& Engineering, University of Michigan, Ann Arbor, MI, USA 48109}%

\author{Kyle Bushick}
\affiliation{ 
Department of Materials Science \& Engineering, University of Michigan, Ann Arbor, MI, USA 48109}%

\author{Andrew McAllister}
\affiliation{ 
Applied Physics Program, University of Michigan, Ann Arbor, MI, USA 48109} %

\author{Woncheol Lee}
\affiliation{ 
Department of Electrical Engineering \& Computer Science, University of Michigan, Ann Arbor, MI, USA 48109}

\author{Chris G. Van de Walle}
\affiliation{ 
Materials Department, University of California, Santa Barbara, CA, USA 93106}

\author{Emmanouil Kioupakis}
\affiliation{ 
Department of Materials Science \& Engineering, University of Michigan, Ann Arbor, MI, USA 48109}%


\begin{abstract}
The quantum efficiency of AlGaN ultraviolet light-emitting diodes (LEDs) declines (droops) at increasing operating powers due to Auger-Meitner recombination (AMR).  Using first-principles density-functional theory, we show that indirect AMR mediated by electron-phonon coupling and alloy disorder can induce bulk $C$ coefficients as large as $\sim10^{-31}$ cm$^6$/s. Furthermore, we find that the confinement of carriers by polarization fields within quantum wells severely relaxes crystal-momentum conservation, which exacerbates the rate of AMR over radiative recombination by an order of magnitude relative to the bulk. This results in a striking decrease in quantum efficiency at high power. Suppressing polarization fields and jointly increasing the well width would greatly mitigate AMR and efficiency droop.
\end{abstract}

\maketitle

AlGaN light-emitting diodes (LEDs) offer remarkable potential as portable and non-toxic light sources in the deep-ultraviolet (UV) wavelengths (200-280 nm), with diverse applications ranging from healthcare to manufacturing.\cite{kneisslEmergenceProspectsDeepultraviolet2019, amano2020UVEmitter2020} Unfortunately, their wall-plug efficiency remains low, particularly at high drive currents. Recent advances in the doping of ultra-wide-band-gap nitride semiconductors have the potential to significantly mitigate extrinsic losses such as poor carrier injection and strong absorption in outcoupling layers, arising from the use of GaN as the contact layer.\cite{pandeyEnhancedDopingEfficiency2019, bagheriHighConductivityGedoped2023, rathkanthiwarHighPconductivityAlGaN2023} Moreover, improvements in the growth and fabrication of AlGaN LEDs continue to yield significant reductions in defect densities and improvements in device performance.\cite{leeMBEGrowthDonor2021, cameronInfluenceThreadingDislocations2022, saifaddinAlGaNDeepUltravioletLightEmitting2020} Therefore, attention must now turn to intrinsic loss mechanisms. These mechanisms are much less well understood due to the difficulty in achieving ideal resonance conditions during the laser excitation of AlGaN LEDs, as well as difficulty in disambiguating internal losses from carrier-transport effects and morphology differences.\cite{frankerlChallengesReliableInternal2019} In this regard, atomistic calculations based on predictive first-principles density-functional theory (DFT) can shed light on the intrinsic performance bottlenecks that limit the efficiency of AlGaN light-emitting devices fabricated with state-of-the-art epitaxy. 

Auger-Meitner recombination\cite{matsakisRenamingProposalAuger2019} (AMR) is a third-order non-radiative recombination mechanism that is suspected to cause efficiency droop, a phenomenon in which the internal quantum efficiency (IQE) declines with increasing current, in AlGaN LEDs.\cite{frankerlChallengesReliableInternal2019, kioupakisFirstprinciplesCalculationsIndirect2015, finnTheoreticalStudyImpact2023, nippertAugerRecombinationAlGaN2018, rudinskyRadiativeAugerRecombination2020, sunEfficiencyDroop2452010, yunAnalysisEfficiencyDroop2015, haoElectricalDeterminationCurrent2017} Electron overflow is another intrinsic mechanism that causes the efficiency to droop,\cite{mondalSuppressionEfficiencyDroop2020} however this can be mitigated with appropriate compositions of the well and barrier regions as well as with the use of an electron-blocking layer. The IQE is defined as $ \text{IQE} = Bn^2/(An + Bn^2 + Cn^3)$, where $n$  is the carrier density, and $An$ , $Bn^2$, and $Cn^3$  are the Shockley-Read-Hall, radiative, and Auger-Meitner recombination terms. Although one may expect that AMR should be negligible in AlGaN due to its ultra-wide band gap,\cite{tamulaitisEfficiencyDroopHighAlcontent2011} experimentally measured $C$ coefficients span the range from $6\times10^{-31}$ to $8\times10^{-30}$ cm$^6$/s,\cite{nippertAugerRecombinationAlGaN2018, haoElectricalDeterminationCurrent2017} and are comparable to the InGaN system.\cite{davidReviewPhysicsRecombinations2019} Moreover, AMR has been proposed to contribute to the degradation of UVB\cite{ruschelCurrentinducedDegradationLifetime2019} and UVC\cite{trivellinPerformanceDegradationCommercial2023} LEDs at high current stress. Scattering processes such as alloy scattering and electron-phonon scattering are known to enhance the AMR rate by enabling transitions to a broader range of final states.\cite{kioupakisIndirectAugerRecombination2011} Moreover, carrier confinement along the $c$-axis in quantum wells introduces new scattering channels that can increase the AMR rate as well.\cite{polkovnikovAugerRecombinationSemiconductor1998} However, the impact of these mechanisms on AMR in AlGaN LEDs is not yet known. 

In this Letter, we explain the origin of the large AMR coefficients in AlGaN UV LEDs. We use first-principles DFT calculations to assess the impact of electron-phonon coupling and alloy disorder on the AMR rate. By explicitly calculating the recombination coefficients in AlGaN quantum wells, we demonstrate that carrier confinement enhances the $C/B$ ratio by approximately one order of magnitude relative to the bulk. This results in a dramatic decrease in the quantum efficiency, particularly at higher carrier densities where efficiency droop is observed. Our results suggest that the detrimental effects can be attributed to carrier confinement induced by the presence of polarization fields. Suppressing those fields and jointly increasing the quantum-well width to lessen the confinement length of carriers can mitigate efficiency droop in AlGaN UV LEDs.

\begin{figure}[ht!]
    \centering
    \includegraphics[width=0.48\textwidth]{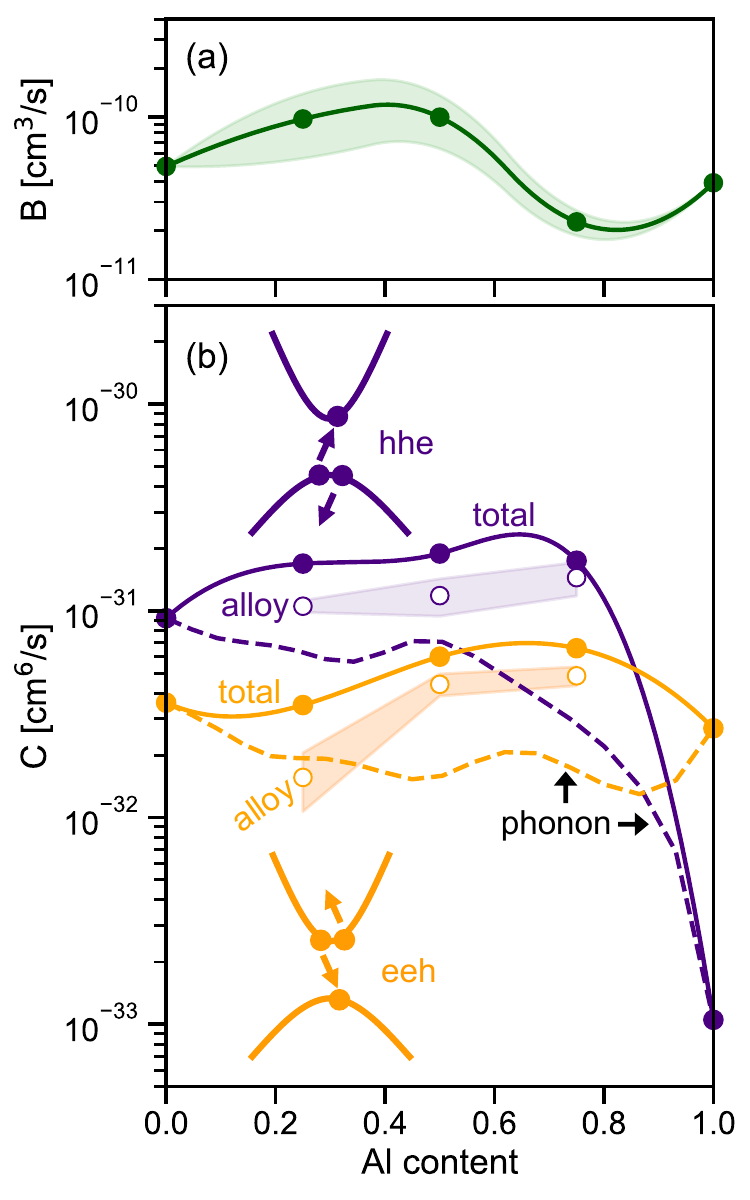}
    \caption{Bulk recombination coefficients of AlGaN alloys for (a) radiative recombination and (b) AMR. The symbols represent averages over eight random configurations, and the shaded regions depict the range within one standard deviation in both directions. The phonon-assisted process is depicted by the dashed lines; refer to the main text for details on our interpolation procedure for the phonon-assisted process. The hole-hole-electron (hhe) $C$ coefficient is given in purple and the electron-electron-hole (eeh) $C$ coefficient is given in orange; “alloy” refers to the alloy-assisted indirect process, “phonon” refers to the phonon-assisted indirect process, and “total” refers to the sum of the alloy- and phonon-assisted processes. The solid lines are spline interpolations solely intended to be a guide to the eye.}
    \label{fig:1}
\end{figure}

To calculate recombination coefficients, we first performed plane-wave DFT calculations using norm-conserving pseudopotentials in the local-density approximation (LDA),\cite{giannozziQuantumESPRESSOExascale2020} and corrected for the band gap through rigid shifts. We calculated the Coulomb matrix elements using first-principles wave functions, and applied density-functional perturbation theory to calculate phonon dispersions and electron-phonon matrix elements in the full Brillouin zone.\cite{baroniPhononsRelatedCrystal2001} We describe our calculations of the $B$ and $C$ coefficients in Section 1 of the Supplementary Material (SM). To manage the combinatorial challenge associated with performing AMR calculations, we followed our established procedure of approximating the cell-periodic part of the Bloch wave functions for the initial states with those at the $\Gamma$ point.\cite{bushickPhononAssistedAugerMeitnerRecombination2023, kioupakisFirstprinciplesCalculationsIndirect2015} For AMR calculations in large supercells, we additionally relied on folding of the Brillouin zone onto the $\Gamma$ point to capture the final states. The decrease in computational cost afforded by this sampling scheme enabled us to sample many random-alloy configurations and significantly increase the supercell size to model realistic heterostructures. We refer the reader to Section 2 of the SM for convergence tests of the AMR coefficients. 

We calculated the recombination coefficients of both bulk alloys and quantum wells. For bulk alloys, we sampled eight 128-atom 4$\times$4$\times$2 special-quasirandom-structure supercells for each composition of 25\%, 50\%, and 75\% Al in AlGaN.\cite{vandewalleEfficientStochasticGeneration2013} Details of these structures and their electronic properties can be found in our previous study on electron-alloy scattering.\cite{pantHighElectronMobility2020} We carefully checked the convergence of our calculations with respect to the supercell size for the alloys (see Figure S9 of the SM), and found the 128-atom system to be sufficiently converged. For the quantum wells, we also sampled five random configurations of 448-atom 4$\times$4$\times$7 and 512-atom 4$\times$4$\times$8 Al$_{0.5}$Ga$_{0.5}$N/AlN quantum wells grown along the $c$-axis. We set the quantum-well widths to five and seven monolayers ($\sim$1 nm and $\sim$1.5 nm, respectively), which are similar to widths used in experimental systems of interest.\cite{frankerlChallengesReliableInternal2019, nippertAugerRecombinationAlGaN2018, frankerlOriginCarrierLocalization2020} We selected nine monolayers ($\sim$2 nm) of AlN as the barrier, and verified that increasing the barrier width to 2.5 nm had only a marginal effect on the recombination coefficients (see Figure S13 of the SM). In all cases, we epitaxially strained the quantum wells to the in-plane lattice constant of AlN, and allowed the internal atomic coordinates and the $c$-axis length to relax. In Figure S2 of the SM, we show that the recombination coefficients are virtually identical between fully relaxed bulk structures and bulk structures that have been epitaxially strained to mimic pseudomorphic strain on AlN substrates.

\begin{figure}[ht!]
    \centering
    \includegraphics[width=0.4\textwidth]{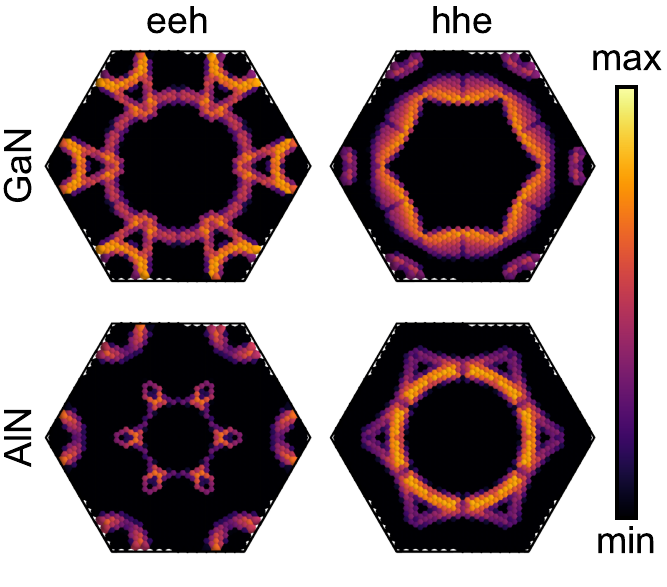}
    \caption{Strength of phonon-assisted AMR as a function of the momentum transferred for the $k_z=0$ cross-section of the Brillouin zone, for a rigidly shifted band gap of 4.7 eV. The dominant AMR transitions in the ultra-wide-band-gap nitrides are short ranged (involve large momentum transfer).}
    \label{fig:2}
\end{figure}

\begin{figure*}[ht!]
    \centering
    \includegraphics[width=0.75\textwidth]{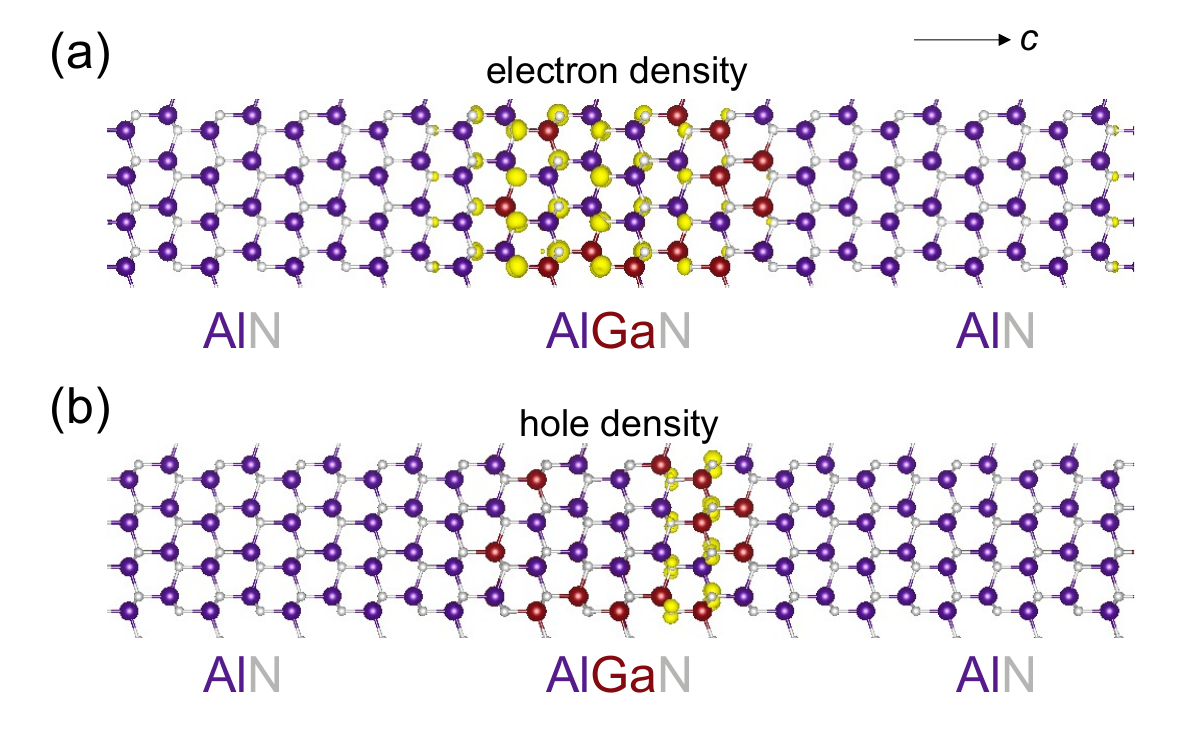}
    \caption{Isosurfaces of the (a) electron and (b) hole charge density at the band extrema, set to 5\% of the maximum density using VESTA,\cite{mommaVESTAThreedimensionalVisualization2011} in the 1.5 nm AlGaN/AlN quantum well. Confinement by the AlN barriers results in the charge density being localized within the AlGaN well. The internal polarization field further induces a separation of charge resulting in the electrons and holes being confined to opposite sides of the well. }
    \label{fig:3}
\end{figure*}

We present our calculations for the radiative and AMR coefficients of bulk GaN, AlN, and AlGaN alloys in Figure 1. We find that the $B$ coefficient increases from $5\times10^{-11}$ cm$^3$/s in GaN to $10^{-10}$ cm$^3$/s in Al$_{0.5}$Ga$_{0.5}$N due to the dependence of the optical-transition matrix elements on the band gap.\cite{rohlfingElectronholeExcitationsOptical2000} The decrease in the $B$ coefficient between 50\% and 75\% Al can be attributed to a corresponding decrease in the wave-function overlap of the optical-transition matrix elements. Our calculated $B$ coefficients are similar in order of magnitude to experimental values by Podlipskas et al., which show an increase in the $B$ coefficient from $10^{-11}$ cm$^3$/s in GaN to $8\times10^{-11}$ cm$^3$/s in Al$_{0.7}$Ga$_{0.3}$N.\cite{podlipskasDependenceRadiativeNonradiative2016} We note that previous empirical calculations of the $B$ coefficient in AlGaN have predicted that the $B$ coefficient decreases with increasing aluminum composition,\cite{dmitrievRateRadiativeRecombination1999} in disagreement with experiment. It is worth noting that the method of Podlipskas et al. prevents them from resolving the carrier-density dependence of the $B$ coefficient and from fully disentangling the radiative and non-radiative contributions to the lifetime. Moreover, we were unable to find experimental measurements of the $B$ coefficient beyond 70\% Al content.

For the AMR process, our calculations show that the hole-hole-electron (hhe) process dominates over the electron-electron-hole (eeh) process due to the greater density of final hole states arising from the six-fold degeneracy of the valence-band manifold. A reversal occurs at very high Al compositions resulting from a lack of energy-conserving final states in the hhe process. For the phonon-assisted process, we calculated the compositional dependence of the $C$ coefficient by performing a gap-dependent interpolation of the $C$ coefficients of GaN and AlN. An analysis of the momentum transferred during phonon-assisted AMR transitions shows that scattering is predominantly short ranged (Figure 2). Furthermore, we find that alloy-assisted AMR is the dominant indirect mechanism for the intermediate compositions where disorder is strongest. For a band gap of 4.7  eV,  which is relevant for UVC germicidal applications, we calculate the total (eeh and hhe) bulk $C$ coefficient to be approximately $2.5\times10^{-31}$ cm$^6$/s.

We now show that quantum confinement exacerbates the $C/B$ ratio. Figure 3 illustrates the simulated structures along the $c$-axis, including the electron and hole charge density at the band extrema. The strong confinement of the charge density within the quantum wells results in mixing of Bloch states with well-defined $k_\parallel$ crystal momenta, and this opens a new channel to break the conservation of crystal momentum in quantum wells, which is important for indirect AMR. Another important feature of the charge density in Figure 3 is the spatial separation of electrons and holes induced by the polarization field, which affects recombination rates by reducing the probability that electrons and holes encounter each other. The effects of quantum confinement and polarization fields are fully captured in Figure 4, which compares the $B$ and $C$ coefficients of Al$_{0.5}$Ga$_{0.5}$N/AlN quantum wells to those of bulk Al$_{0.5}$Ga$_{0.5}$N alloys as a function of the band gap. The macroscopic polarization field decreases the $B$ coefficient by a factor of 1.8 for the 1 nm quantum well and 7.6 for the 1.5 nm quantum well. The $C$ coefficient is also affected by the polarization field, however this reduction is compensated by the additional transitions enabled by the heterostructure-induced symmetry reduction. For the 1 nm quantum well, the eeh and hhe coefficients increase by a factor of 2.9 and 4.5, respectively. On the other hand, for the 1.5 nm quantum well, the eeh coefficient decreases by a factor of 2.8 while the hhe coefficient increases by a factor of 1.14 relative to bulk. 

\begin{figure}
    \centering
    \includegraphics[width=0.4\textwidth]{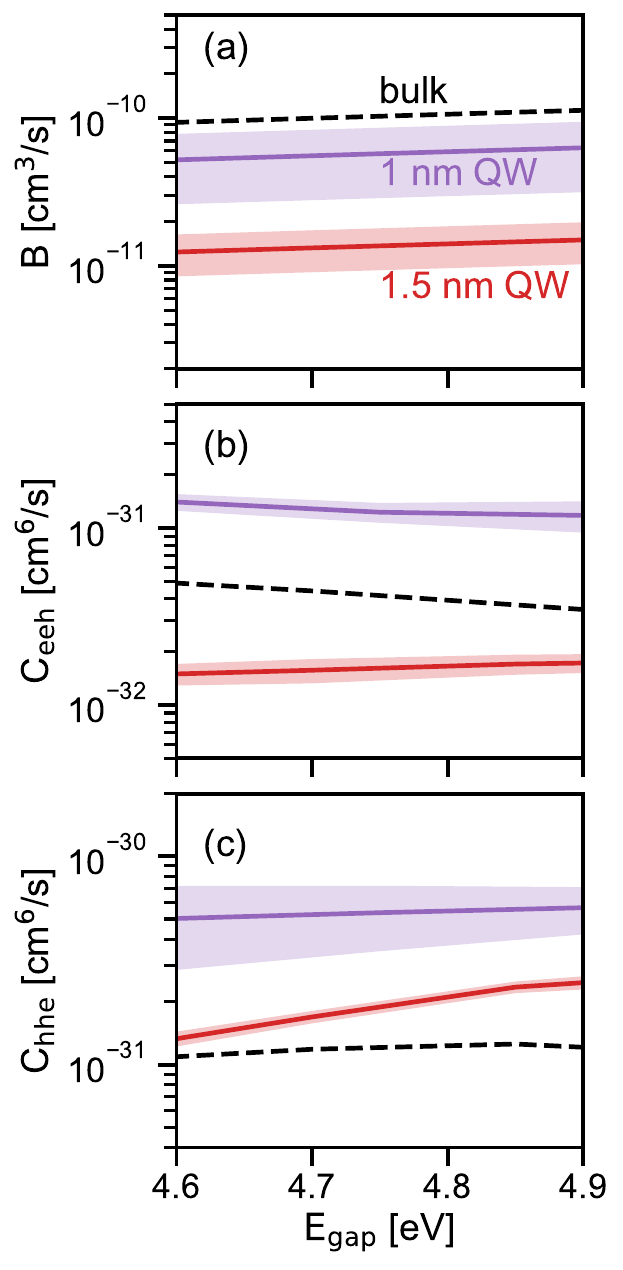}
    \caption{Comparison of recombination coefficients in bulk vs quantum wells for (a) radiative recombination, (b) eeh AMR, and (c) hhe AMR. The solid line is an average over five random configurations, and the shaded region depicts the range within one standard deviation in both directions. Carrier confinement exacerbates the $C/B$ ratio relative to bulk.}
    \label{fig:4}
\end{figure}

\begin{figure}[htp]
    \centering
    \includegraphics[width=0.4\textwidth]{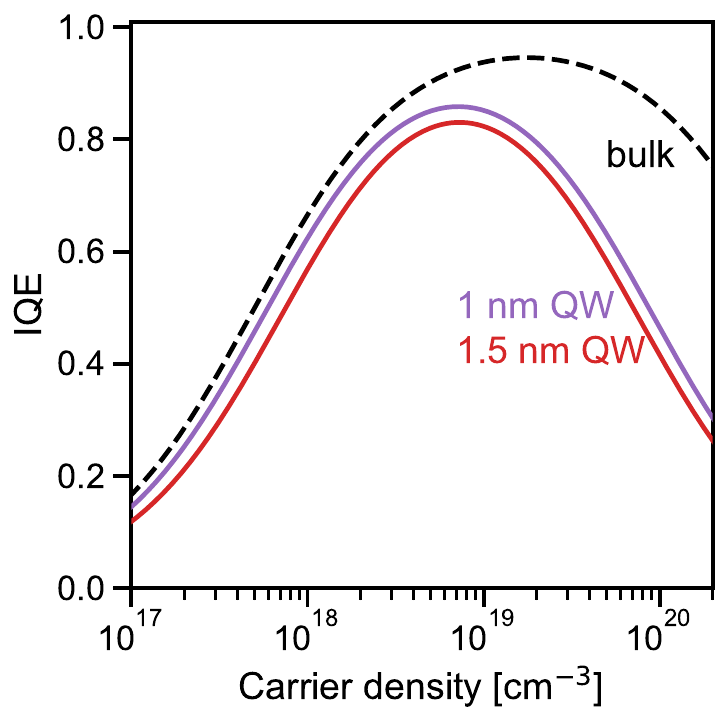}
    \caption{Internal quantum efficiency of bulk Al$_{0.5}$Ga$_{0.5}$N alloys compared to Al$_{0.5}$Ga$_{0.5}$N/AlN quantum wells with quantum-well widths 1 nm and 1.5 nm. Efficiency droop is significantly aggravated for quantum wells compared to bulk materials.}
    \label{fig:5}
\end{figure}

Note that we have omitted the phonon-assisted contributions in this analysis as they are prohibitively expensive to calculate in low-symmetry heterostructures, and alloy-assisted AMR is dominant in the composition range of interest. Nevertheless, since the phonon-assisted AMR transitions are predominantly short ranged (involve large phonon-momentum transfer $q$), the bulk values are a good starting approximation. We expect the mixing of states with well-defined $k_\parallel$ to also exacerbate the phonon-assisted $C$ coefficient relative to bulk, therefore our $C$ coefficients for the quantum wells serve as lower-bound estimates. For a band gap of 4.7 eV, we estimate the sum of the eeh and hhe $C$ coefficients to range between $5\times10^{-31}$ to $9\times10^{-31}$ cm$^6$/s for the 1 nm quantum well, and between $1.7\times10^{-31}$ cm$^6$/s to $2\times10^{-31}$ cm$^6$/s for the 1.5 nm quantum well.

The difference in the $C/B$ ratio in quantum wells leads to a dramatic drop in the IQE compared to the bulk, particularly in the efficiency-droop regime. Figure 5 shows that the IQE is significantly lower in quantum wells than in the bulk, with the difference becoming larger at increasing carrier densities. In this analysis, we have used the $B$ and $C$ coefficients from Figure 4 assuming a band gap of 4.7 eV, and set the bulk $A$ coefficient to $5\times10^7$/s, which is the same order of magnitude as experiment.\cite{nippertAugerRecombinationAlGaN2018} We describe our treatment of the $A$ coefficient in Section 1 of the SM; our analysis is not sensitive to the precise value of $A$. Despite large differences in the $B$ and $C$ coefficients between the 1 nm and the 1.5 nm quantum well, we find their ratio to be remarkably similar, with $C/B$ values of $8.6\times10^{-21}$ cm$^3$ and $1.4\times10^{-20}$ cm$^3$ for the 1 nm and 1.5 nm quantum wells. This represents nearly one order of magnitude increase of the $C/B$ ratio in quantum wells relative to the bulk value of $1.6\times10^{-21}$ cm$^3$. In general, we expect our conclusions regarding the IQE to be valid for quantum wells structures of different geometries because the dependence of $B$ and $C$ on the well width cancels out in their ratio. The net outcome is that the $C/B$ ratio is significantly larger in quantum wells than in bulk.  

Our results clearly indicate that overcoming the intrinsic AMR limit of AlGaN quantum wells is possible by changing the quantum-well design. In contrast to conventional III-V quantum wells, the potential profile in polar AlGaN quantum wells is triangular, and it is the polarization field that confines the carrier. Therefore, increasing the quantum-well width does not lessen the degree of carrier confinement unless the polarization field is also removed (see Figure S14 of the SM for a schematic illustration of this concept). Suppressing the polarization fields by growing devices on non-polar planes,\cite{dinhAluminiumIncorporationPolar2019} by doping the barriers as demonstrated already in visible emitters\cite{chowImpactDopedBarriers2022} or by growing ultra-thick quantum wells\cite{uhligTransitionQuantumConfinement2023} could be promising methods to increase the IQE. The width of such polarization-free heterostructures should be large to minimize the impact of quantum confinement on AMR. Digital alloys based on atomically thin superlattices of AlN and GaN offer another possibility to further reduce the AMR rate because these structures are free of alloy disorder, thus they remove the alloy-scattering channel for AMR transitions.\cite{bayerlDeepUltravioletEmission2016} Our previous analysis of these structures suggests that they can be grown as thick layers on AlN, and growth on $m$-plane substrates or doping of the barriers would enable these structures to be free from polarization fields as well.\cite{pantIncreasingMobilityPowerelectronics2022} Ultra-thin GaN emitters embedded in AlN hosts have already been demonstrated experimentally,\cite{islamMBEgrown2322702017,nichollsHighPerformanceHigh2023} however the width of the AlN layers needs to further decrease to the monolayer limit in order for these structures to behave as true digital alloys with charge densities that are extended throughout the superlattice rather than as ultra-thin quantum wells with the charge density confined to the GaN layers. 

In conclusion, we have calculated the radiative $B$ and AMR $C$ coefficients in bulk AlGaN alloys and quantum wells. We identified alloy scattering and electron-phonon coupling as mechanisms that exacerbate the AMR rate in bulk, and demonstrated that symmetry reduction in $c$-plane quantum wells further increases the $C/B$ ratio, leading to a significant decrease in the IQE. Overcoming the intrinsic AMR limit is possible by eliminating alloy disorder with the use of atomically thin superlattices, as well as minimizing carrier confinement by eliminating polarization fields and increasing the quantum-well width.

\vspace{-5mm}
\section*{Supplementary Material}
\vspace{-7mm}
Refer to the SM for further technical details on the density-functional and density-functional-perturbation theory calculations, explanations of our calculations of the recombination coefficients, and a comparison of how epitaxial strain affects the bulk recombination coefficients. The SM also contains convergence tests for the phonon-assisted AMR coefficient for AlN and GaN as well as convergence tests for indirect AMR rates calculated for supercell bulk-alloy and quantum-well structures. Finally, the SM contains schematic illustrations of the qualitative differences in carrier confinement between polar and polarization-free quantum wells. 
\vspace{-5mm}
\section*{Acknowledgements}
\vspace{-7mm}
This work was primarily supported as part of the Computational Materials Sciences Program funded by the U.S. Department of Energy (DOE) Office of Science, Basic Energy Sciences (BES) under Award No. DE-SC0020129 (method development and calculations of radiative and AMR recombination rates, discussion of AlGaN materials; N.P., K.B., A.M., W.L., E.K.) and by the U.S. DOE Office of Science, BES under Award No. DE-SC0010689 (discussion of AlGaN devices, C.G.V.d.W.). Computational resources were provided by the National Energy Research Scientific Computing Center, a DOE Office of Science User Facility, supported under Contract No. DE-AC02-05CH11231. N. P. acknowledges the support of the Natural Sciences \& Engineering Research Council of Canada Postgraduate Scholarship. K. B. acknowledges the support of the U.S. DOE, Office of Science, Office of Advanced Scientific Computing Research, Department of Energy Computational Science Graduate Fellowship under Award No. DE-SC0020347. A. M. acknowledges support from the NSF Graduate Research Fellowship Program (No. DGE-1256260). W. L. was partially supported by the Kwanjeong Educational Foundation Scholarship.
\vspace{-5mm}
\section*{Data availability}
\vspace{-7mm}
The data that support the findings of this study are available from the corresponding authors upon reasonable request. 
\vspace{-5mm}
\section*{References}
\vspace{-7mm}

\bibliography{AlGaN_droop_biblatex}

\end{document}